# Tuning Phononic and Electronic Contributions of Thermoelectric in defected S-Shape Graphene Nanoribbons

M.Amir Bazrafshan, Farhad Khoeini*

*Department of Physics, University of Zanjan, P.O. Box 45195-313, Zanjan, Iran*

**Abstract**

Thermoelectrics as a way to use waste heat, is essential in electronic industries, but its low performance at operational temperatures makes it inappropriate in practical applications. Tailoring graphene can change its properties. In this work, we are interested in studying the transport properties of S-shape graphene structures with the single vacancy (SV) and double vacancy (DV) models. The structures are composed of a chiral part, which is an armchair graphene nanoribbon, and two zigzag graphene ribbons. We investigate the changes in the figure of merit by means of the Seebeck coefficient, electronic conductance, and electronic and phononic conductances with the vacancies in different device sizes. The transport properties of the system are studied by using the non-equilibrium Green's function method, so that the related Hamiltonians (dynamical matrices) are obtained from the tight-binding (force constant) model. The maximum figure of merit (ZT) obtains for the DVs in all lengths. Physical properties of such a system can be tuned by controlling various parameters such as the location and the type of the defects, and the device size. Our findings show that lengthening the structure can reduce phononic contribution, and single vacancies than double vacancies can better distinguish between electronic thermal conductance behavior and electronic conductance one. Namely, vacancy engineering can significantly increase thermoelectric performance. In the large devices, the SVs can increase the ZT up to 2.5 times.

## 1. Introduction

Modern life is entangled with computers, and transistors are the heart of the computer processors. As the number of transistors in an electronic chip rises, the power density also increases, decreasing the integrated circuits performance reliability [1]. In electronic devices, maintaining the temperature in an appropriate range is very important [2]. Lowering their temperature can be achieved by dissipating the heat or converting it into another form of energy and using it. For heat dissipation, high thermal conductance is essential, and to convert heat into another form of energy, such as electricity, thermoelectric is one of the solutions, but it requires low thermal conductance. Researchers are trying to find the best material for thermoelectrics [3]. They face some challenges. For example, electron scattering in junctions produces heat, so the material must be a good electronic conductor to produce less heat, and to enhance the thermoelectric performance, the temperature difference should be maintained [4].

Many materials can be fabricated, thanks to nowadays facilities. But this is not enough if someone wants to synthesize materials for a specific goal with trial and error. Theoretical approaches can provide an easy way to discover the underlying factors involved in the properties of materials. Based on theoretical approaches, researchers are trying to find the best material for this purpose. Theoretically, using an algorithm, in an edge defected long ZGNR, ZT is reported four in Ref. [5] at room temperature in a 4 μm long ZGNR. Other works also reported a figure of merit of six for narrow AGNRs at room temperature [6], 0.88 for a bent structure with 24 pores [7] at 500 K, up to three in a graphene structure with different percentage of carbon isotopes doping, and vacancies [8], ZT>2 for a 2 μm long ZGNR with extended line defect [9], and ZT ≃ 11 at 77 K for ZGNR based devices with a length of >1 μm containing two special nonperiodic nanopores with different diameters [10]. As mentioned in Ref. [11], for long GNRs, the electron-phonon interaction may not be neglected since it might be significant in long GNRs, though we neglect it since our structures are small enough [12]. Besides, the Umklapp scattering is not considered for the same reason [13].

Graphene is the first 2D successfully synthesized material [14]. It has the highest thermal conductance until now [15], which makes it the best candidate to conduct heat for dissipation applications [11,16]. However, as the dimensions decrease, quantum confinement effects become important [17], which can help maximize ZT [18] by manipulating physical properties. This can be accompanied by phonon scattering due to nanostructure boundaries [19]. In 2D materials, especially hexagonal structures, the edge geometry of a ribbon provides a degree of freedom to tailor its physical properties [16,20,21]. Introducing defects, doping, and applying mechanical strain can also alter the physical properties of graphene [22–25].

Since the shape and geometry of the nanodevices are important in tuning physical properties at the nanoscale, we are interested in studying the S-Shape graphene structures with three different lengths. S-Shape structure is a mix-up of zigzag and armchair edge geometries, which can help to tune physical properties. In an S-Shape graphene nanoribbon (GNR), electronic contributions can be significantly altered due to the quantum confinement and edge effects [26]. In this work, the temperature is considered 350 K, close to what is to be controlled in processor units [27,28]. To evaluate the thermoelectric performance, the figure of merit, a dimensionless parameter, is investigated. The figure of merit can be calculated as $ZT(\mu, T) = \frac{gS^2}{\kappa_e + \kappa_{ph}} T$, with $g$ as electronic conductance, $S$ as Seebeck coefficient, $\kappa_e$ as electronic thermal conductance, $\kappa_{ph}$ as phononic thermal conductance, and $T$ as absolute temperature. These parameters are individually plotted for each of the studied structures.

The two experimentally observed vacancies, single vacancy (SV) and divacancy (DV) [29], are introduced, and their impact on both electronic and phononic contributions related to thermoelectric performance is studied.

We have used the non-equilibrium Green's function (NEGF) method to calculate the interested quantities. Hamiltonians are obtained from the tight-binding (TB) approach by considering up to third nearest-neighbor (3NN) interactions. To be more accurate [30], overlap integrals are taken into account. For phononic thermal conductance, force tensor matrices obtained via the force constant (FC) model by considering up to 4NN.

Vacancies are introduced and named as indicated in figure 1, e.g., a single vacancy located at the eleventh atomic position in the armchair direction and the fifteenth one in the zigzag direction is identified by SV-11-15; the number of atomic positions is also presented for each direction.

The article is arranged as follows; in the next section, we will describe the model, with a brief introduction on the TB, and FC formulations. Results and discussion are in section three. In the last section, we conclude our study.

## 2. Model and Method

In this section, we describe a system consisting of left and right contacts and a central device connected to them.

To start, a schematic of the structure of system is presented in figure 1. The system is divided into three parts with black boxes; the device section is an S-Shape graphene structure, the right and left contacts are two semi-infinite ZGNRs with a width of 12 atoms. The grey dashed boxes show the unit cells in contacts. Also, the first, second, and third nearest neighbors are displayed with concentric circles in the left contact, so that magenta dashed circle shows the first nearest-neighbor domain, the cyan dashed circle shows the 2NN domain, and red dashed circle indicates the 3NN domain. Vacancies are identified by their position in the armchair and the zigzag edge geometries. The numbers on the left and the bottom of the device section are for easy identification of those vacancies. Three examples of how to identify vacancies are shown. Vacancies are identified with the general form of VT-m-n-or, in which the VT is the vacancy type, here it can be SV or DV, m and n are atomic positions in the armchair and zigzag directions, respectively, and the last part indicates respective orientation to the nanoribbon axis. For single vacancies, it is omitted, but for divacancies, the relative orientation of the hypothetical line between two removed atoms determines the last part. If a DV is perpendicular to the ribbon axis (or parallel to the ribbon width), it is indicated with 'pr', it is indicated with 'or'. To be more precise, the cyan box in figure 1 indicates a divacancy DV-7-8-or, in which its first atom (as numbers, from left to right) is in the seventh atomic position in the armchair direction, and the eighth atomic position in the zigzag direction, this divacancy is oriented respect to the ribbon axis which is indicated by "or" in the name of the DV. The direction, in which vacancies move in the structure is marked by the green arrow. The dashed lines are bonds that are affected throughout the study. The hatched area shows the zone where the vacancies are introduced.

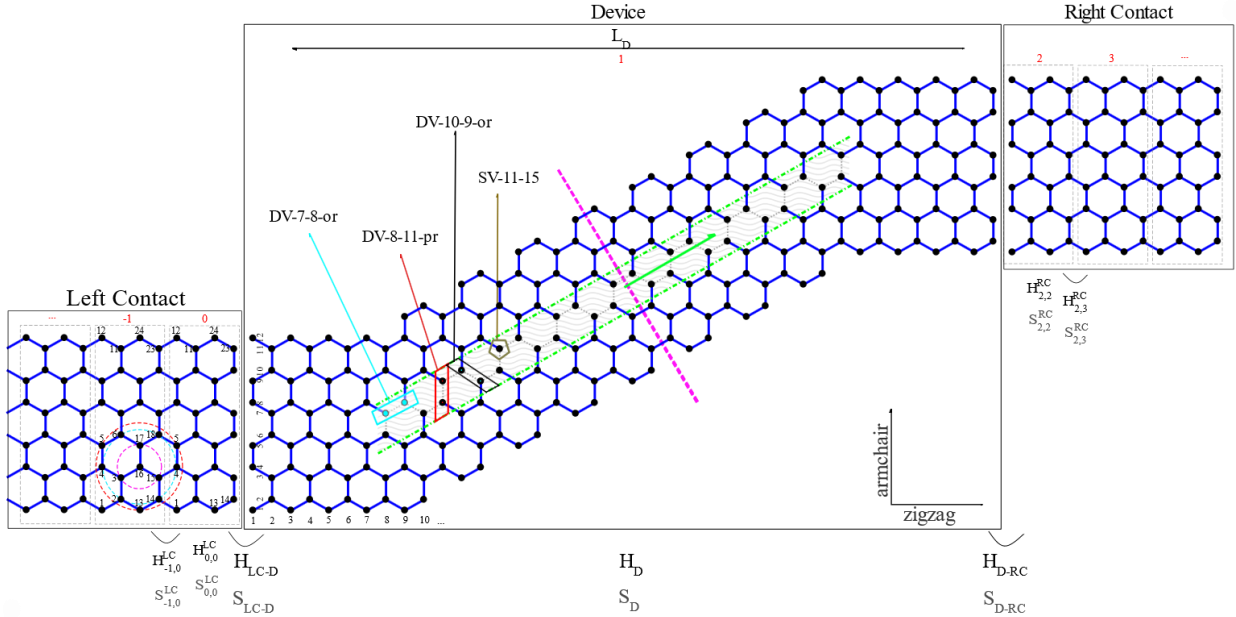

**Figure 1** A schematic of the model implemented in the NEGF method with all Hamiltonians and overlap matrices. Vacancies are relocated in the hatched area, and four vacancies with their names are shown. The thick purple dashed line indicates the middle of the device section. Three concentric circles show the first, second, and third nearest-neighbor domains, respectively, by magenta, cyan, and red colors for a selected atom. Vacancies were considered in different places in the hatched area (the green arrow shows this direction). Numbers in the device were used for this purpose as described in the text.

To employ the NEGF method for electronic and phononic contributions, matrices that describe electron and phonon energies and their interaction with $n^{th}$ nearest neighbors, are essential. To form matrices for electrons (i.e., Hamiltonians) in the tight-binding approach, the unit cell should be defined (as depicted in figure 1 with dashed gray rectangles). In the non-orthogonal tight-binding approach, the Hamiltonian of the system, its elements, and the elements of overlap matrix are as [7]:

$$H = \sum_i \varepsilon_i |i\rangle\langle i| + \sum_{<i,j\in \overline{3}NN>} (-t_{i,j}|i\rangle\langle j|+h.c), \quad H_{i,j} = \langle i|H|j\rangle, \tag{1a}$$

$$S_{i,j} = \langle i|j\rangle, \tag{1b}$$

where $\varepsilon_i$ is the on-site energy and $t_{i,j}$, and $S_{i,j}$ are the interatomic and overlap parameters, respectively. There are several sub and superscripts that LC means the left contact, RC means the right contact, and D is the device. As indicated in figure 1, $H_{0,0}^{LC}$ is the Hamiltonian of the unit cell 0 in the left contact, and $H_{-1,0}^{LC}$ is the coupling Hamiltonian between the unit cell number -1 and 0 in the left contact.

Hopping and overlap parameters are presented in table 1 as reported in Ref. [31]. The electronic energy dispersion for a periodic system, like the left contact, can be obtained by solving the eigenvalue problem [32]:

$$\det[H_k - ES_k] = 0, \tag{2}$$

where $H_k$ and $S_k$ are given by:

$$H_k = H_{0,0}^{LC} + H_{-1,0}^{LC}\exp(-ika) + \left(H_{-1,0}^{LC}\exp(-ika)\right)^\dagger, \tag{3a}$$

$$S_k = S_{0,0}^{LC} + S_{-1,0}^{LC}\exp(-ika) + \left(S_{-1,0}^{LC}\exp(-ika)\right)^\dagger, \tag{3b}$$

in which $k$ and $a$ are the wave vector and the lattice constant, respectively. The transmission probability (for electrons $T_e$, and for phonons it is shown by $T_{ph}$) can be calculated using Green's function method [33]; details are provided in the supplementary materials.

By having transmission probability, one can calculate the electronic conductance $g(\mu, T)$, the Seebeck coefficient $S(\mu, T)$, and the electronic thermal conductance $\kappa_e(\mu, T)$ as [11,34]:

$$g(\mu, T) = e^2 L_0(\mu, T), \tag{4a}$$

$$S(\mu, T) = \frac{1}{eT}\frac{L_1(\mu, T)}{L_0(\mu, T)}, \tag{4b}$$

$$\kappa_e(\mu, T) = \frac{1}{T}\left[L_2(\mu, T) - \frac{L_1^2(\mu, T)}{L_0(\mu, T)}\right], \tag{4c}$$

here $e$ is the elementary charge, and $L_n$ is given by:

$$L_n(\mu, T) = -\frac{2}{h}\int_{-\infty}^{\infty} T_e(E)\frac{(E-\mu)^n}{k_B T}\frac{\exp\frac{(E-\mu)}{k_B T}}{\left(\exp\frac{(E-\mu)}{k_B T}+1\right)^2}dE, \tag{5}$$

which its numerical form is as follows:

$$L_n(\mu, T) = -\frac{2}{h}\sum_E T_e(E)\frac{(E-\mu)^n}{k_B T}\frac{\exp\frac{(E-\mu)}{k_B T}}{\left(\exp\frac{(E-\mu)}{k_B T}+1\right)^2}\Delta E, \tag{6}$$

with $h$ as plank constant and $k_B$ as the Boltzmann constant. This is the discrete form of integral. The summation is over the whole energy range. By considering $l$ as total steps, the integration element for numerical integration in the rectangular method is $\Delta E = (E_{\text{final step}} - E_{\text{initial step}})/(l-1)$.

The secular equation for phonons, which derives from Newton's second law, is:

$$DU = \omega^2 U, \tag{7}$$

in which, $U$ is the matrix containing the vibrational amplitude of all atoms, $\omega$ is the angular frequency, and $D$ is the dynamical matrix:

$$D = [D_{i,j}^{3\times 3}] = \begin{bmatrix} -\dfrac{K_{i,j}}{\sqrt{M_i M_j}} & \text{for } j \neq i \\ \sum\limits_{j\neq i} \dfrac{K_{i,j}}{M_i} & \text{for } j = i \end{bmatrix}. \tag{8}$$

where $M_i$ is the mass of the $i^{\text{th}}$ atom, and $K_{i,j}$ represents $3 \times 3$ force tensor between the $i^{\text{th}}$ and $j^{\text{th}}$ atoms:

$$K_{i,j} = U^{-1}(\theta_{i,j}) K_{i,j}^0 U(\theta_{i,j}), \tag{9}$$

with $\theta_{ij}$ as the angle between the $i^{\text{th}}$ and the $j^{\text{th}}$ atom. The unitary matrix $U(\theta_{i,j})$ is defined by the rotation matrix in a plane as:

$$U(\theta_{i,j}) = \begin{pmatrix} \cos\theta_{i,j} & \sin\theta_{i,j} & 0 \\ -\sin\theta_{i,j} & \cos\theta_{i,j} & 0 \\ 0 & 0 & 1 \end{pmatrix}, \tag{10}$$

also $K_{i,j}^0$ is given by:

$$K_{i,j}^0 = \begin{pmatrix} \varphi_r & 0 & 0 \\ 0 & \varphi_{t_i} & 0 \\ 0 & 0 & \varphi_{t_o} \end{pmatrix}, \tag{11}$$

where $\varphi_r, \varphi_{t_i}, \varphi_{t_o}$ are force constant parameters in the radial, in-plane, and out of plain directions of the $j^{th}$ atom, respectively. To be more clear about these matrices, e.g., for the $D_D$, which represents the dynamical matrix of the device section, regarding Eq. 7, and to write what each atom feels (or when $i = j$), one must consider all 4 NN effects in the summation, including atoms in the neighboring unit cells, i.e., $D_{D_{i,j}} = \sum_i \sum_{j \in 4NN} K_{i^D,j^D} + K_{i^D,j^{RC}} + K_{i^D,j^{LC}}$. For coupling terms like the $D_{D-RC}$ elements, the interaction between the first atom of the device and the first atom of the right neighbor (i.e., diagonal elements) is already accounted, so one can safely set this to zero. Force constants [35], and other essential parameters are presented in Table 1.

Phononic band structure can be obtained by solving the following eigenvalue problem [36]:

$$\left(\sum_j K_{i,j} - \omega^2(\text{k})\text{I}\right)\delta_{i,j} - \sum_j K_{i,j} e^{i\text{k}.\Delta\text{r}_{i,j}} = 0 \tag{12}$$

with $\Delta \text{r}_{i,j} = \text{r}_i - \text{r}_j$ as the distance between the $i^{\text{th}}$ and $j^{\text{th}}$ atoms, and k as the wave vector. To calculate the phononic density of states or vDOS, with v as vibrational, one can use $\text{vDOS} = -\frac{2\omega}{\pi}\text{Im}[\text{Trace}(G(\omega))]$, within Green's function method [37,38], or by using Gaussian smearing of the Dirac Delta [7]:

$$\text{vDOS} = \sum_n \sum_{k \in BZ} \frac{1}{\eta\sqrt{\pi}} e^{-\frac{(\omega-\omega_n(k))^2}{\eta^2}}, \tag{13}$$

with $n$ as band index of the phonon, $k$ as wave vector, and $\eta$ is a small positive number. Note that if the frequency is $\mathcal{O} \times 10^{14}$, then $\eta$ should be a small positive number respect to this order, i.e., $\eta = \mathcal{O} \times 10^{10}$. By having the transmission function, one can obtain vibrational conductance as [11]:

$$\kappa_{\text{ph}}(T) = \frac{1}{8\pi k_B T^2} \int_0^\infty \hbar^2 \omega^2 \frac{T_{\text{ph}}(\omega)}{\sinh^2\left(\frac{\hbar\omega}{2k_B T}\right)} d\omega. \tag{14}$$

When $L_D$ is much shorter than the phonon mean free path (MFP), phononic transport is considered ballistic [39]. The MFP is also a function of the width of the GNR [24,39]. Here, we assume phononic transport is ballistic, because the devices are small enough, and the width of the system is relatively small.

In the next section, we investigate the length of the structure and defect location effect on the electronic and phononic contributions in the ZT formula for the studied structures.

Table 1. Parameters involved in the TB, FC and NEGF methods.

| Parameter (Symbol) | Value (Unit) |
|---|---|
| Carbon on-site energy ($\varepsilon_c$) | -0.187 (eV) |
| Carbon-carbon bond length ($a_{C-C}$) | 1.42 (Å) |
| Mass of the carbon atom ($M_C$) | $1.994 \times 10^{-26}$ (Kg) |
| **Hopping parameters (eV)** | |
| $t_1$ | 2.756 |
| $t_2$ | 0.071 |
| $t_3$ | 0.38 |
| **Overlap parameters (eV)** | |
| $S_1$ | 0.093 |
| $S_2$ | 0.079 |
| $S_3$ | 0.070 |
| **Force constants (N/m)** | |
| $\varphi_r^1$ | 409.8 |
| $\varphi_{t_i}^1$ | 145.0 |
| $\varphi_{t_o}^1$ | 98.9 |
| $\varphi_r^2$ | 74.2 |
| $\varphi_{t_i}^2$ | -40.8 |
| $\varphi_{t_o}^2$ | -8.5 |
| $\varphi_r^3$ | -33.2 |
| $\varphi_{t_i}^3$ | 50.1 |
| $\varphi_{t_o}^3$ | 5.8 |
| $\varphi_r^4$ | 6.5 |
| $\varphi_{t_i}^4$ | 5.5 |
| $\varphi_{t_o}^4$ | -5.2 |

## 3. Results and discussion

We have studied three structures with different lengths to find out what contributions are played a major role in determining thermoelectric performance in S-Shape graphene ribbons and seeking any meaningful vacancy place impact on it. First, an S-shape device with $L_D \approx 34.43$ Å is studied. For this configuration, we have moved vacancies in the shaded area between two green dash-dotted lines in the green arrow direction indicated in figure 1. All terms in the ZT formula are plotted by taking the ratio between the defected and pristine values. We call this hereafter *the ratio*. Also, we will mention $\kappa_e$ and g as *electronic terms*.

The 30° chiral part of the system is a 10-AGNR with a finite length. Electronic band structures and phononic dispersions plotted in figure 2 are for an infinite length of GNRs. A 10-AGNR with infinite length is a semiconductor, as evidenced by the electronic band structure plotted in figure 2(a), with an energy gap ~1.1 eV. The Fermi energy is zero. The band structure is not symmetric respect to the Fermi energy, which is because of the inclusion of the 3NN (with overlap) TB model.

The left part of panels (b) and (d) in figure 2 show the phononic dispersion and the vDOS for the 10-AGNR and the 12-ZGNR, respectively. The acoustic bands are located at low-frequencies and they usually possess much higher group velocities compared to those of the optical ones, so they contribute mostly to the thermal transport. Therefore, low-frequency bands play a dominant role in thermal conductance. Comparing low-frequency phonon bands of the AGNR with ZGNR, shows that phononic bands in the AGNR are less dispersive respect to the ZGNR, therefore, since the phononic transmission coefficient is equal to the sum of phonon modes, dispersive phonon bands lead to larger values of the transmission coefficients. The above discussion suggests that a ZGNR can act as good thermal conductors, while the AGNR is a better candidate for thermoelectric applications. Figure 2(c) shows electronic dispersion for the 12-ZGNR, with no energy gap. Therefore, it is a meal, which is consistent with the fact that all zigzag nanoribbons are metal in simple TB model [40,41]. Metallicity can strongly reduce the ZT value in ZGNRs [42]. The vibrational density of states, vDOS, are plotted in the right part of panels (b) and (d) in figure 2. Also, for low-frequencies, the AGNR has the lower vDOS values than the ZGNR (check the black dotted line on figure 2(e)). The thermal conductance of the AGNR is lower than that of the ZGNR, which is due to the lower phonon density of states at low-frequencies. The above results are consistent with Refs. [31,43,44].

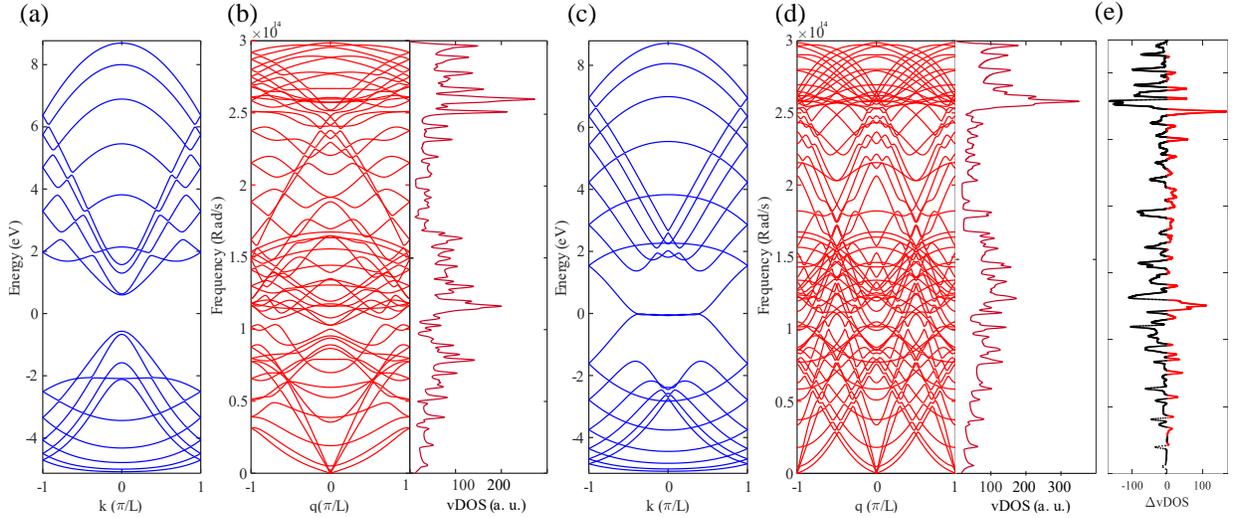

**Figure 2** (a) Electronic band structure for the 10-AGNR, (b) phononic band structure and the vibrational density of states for the 10-AGNR, (c) electronic dispersion, and (d) phononic dispersion plot with the vDOS for the 12-ZGNR, (e) the difference between the vDOSs of the 10-AGNR and 12-ZGNR. The black dotted line shows that the vDOS for the 10-AGNR is smaller than that of the 12-ZGNR.

As shown in figure 3, all electronic and phononic terms, together with the Seebeck coefficient and ZT, show a symmetric behavior respect to the middle of the device (purple dashed line in figure 1). Figure 3(a) shows the maximum ZT (magenta squares) as a function of SV locations. The chemical potential for each $ZT_{max}$ is indicated by numbers near the symbols. In this spectrum, the maximum figure of merit, $\simeq 0.12$, is for SV-10-12 (or SV-13-19) at $\mu = -1.54$ eV.

According to figure 3(b), SVs can alter the $ZT_{max}$ (magenta circles) up to 1.18 times, and Seebeck coefficient square (black stars) up to 4 times greater than the pristine structure. As shown in figure 3(c), by increasing $\kappa_e$ and g and reducing $S^2$, vacancies can also degrade the ZT. Electronic and phononic terms can increase or decrease by varying the place of the defect. Inclusion of an SV, decreases the phononic term, and changes it to ~0.8 of its pristine case, and the electronic ones fluctuate between 1.5 and 0.1 times of its defect-free structure. The $\kappa_e$ is shown by the square symbol, g by the circle symbol, and $\kappa_{ph}$ by stars. The SV-9-11 reduces the ZT to 0.71 times of the pristine structure. This reduction is because of increasing $\kappa_e$ and g to 1.48 times, accompanied by reducing $S^2$ to 0.48 times compared to the pristine case.

The highest ZT value with a DV, $\simeq 0.21$, in $L_D \approx 34.43$ Å, is achieved for the DV-6-8-pr (DV-16-23-pr) (figure 3(d)) at $\mu = -1.52$ eV. In the presence of the DVs, $S^2$ and the $ZT_{max}$ are about 4.3, and 2.1 times greater than the pristine structure (figure 3(e)), respectively. Also, we should mention that the ZT for this vacancy can decrease by 0.52 times. The behavior of electronic and phononic terms, $\kappa_e$, g, and $\kappa_{ph}$ are shown in figure 3(f). A meaningful change in the figure of merit occurs whenever electronic terms take apart from each other. Since double vacancies can reduce available conduction channels for phonons, they affect $\kappa_{ph}$ more than single vacancies, as evidenced in figure 3(f). By recalling the ZT formula, higher g, i.e., lower suppression on g in comparison to $\kappa_e$, can help to achieve the higher ZT and vice versa. In this length, all terms are affected comparably large in the case of the DVs than the SVs.

One can see that the SV-7-8 and SV-16-23 are located in the chiral section of the system. Figure 3(c) shows that electronic terms between these two are symmetric with respect to the center of the device section, and out of that, their behavior is different.

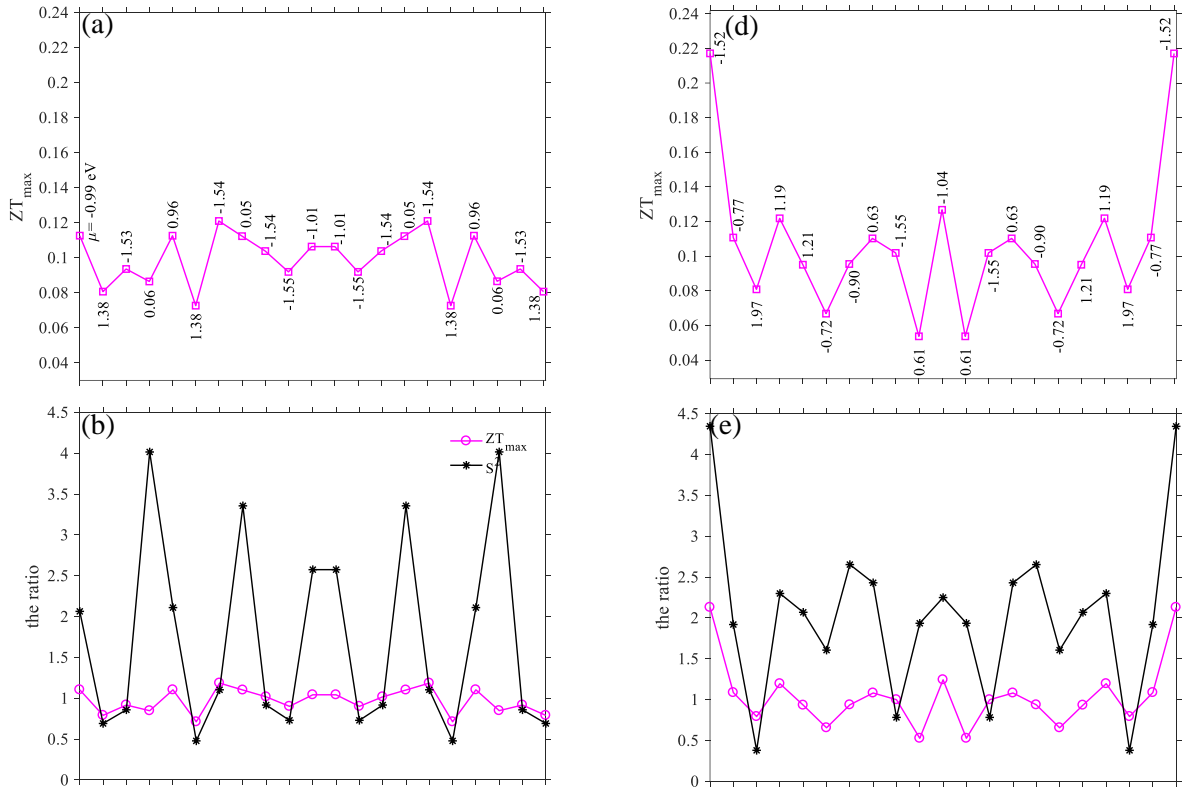

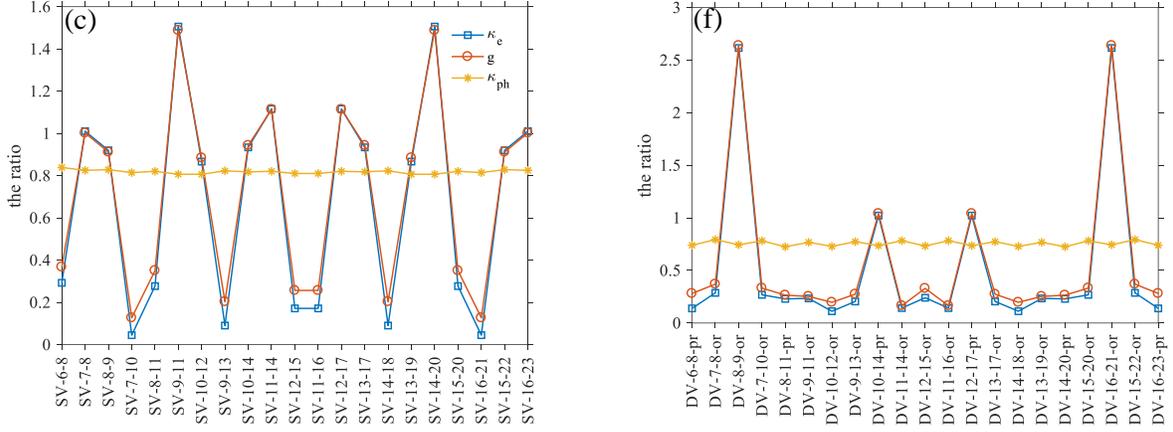

**Figure 3** (a) Maximum figure of merit for different SV locations together with the corresponding μ, (b) the ratio of $S^2$ and $ZT_{max}$, (c) *the ratio* of electronic and phononic terms as a function of various SV locations, (d) the $ZT_{max}$ with the corresponding chemical potential as a function of different DV locations, (e) the ratio of the Seebeck square and the maximum of ZT, and (f) *the ratio* of electronic and phononic terms versus the DV locations. The device length is $L_D \approx 34.43$ Å. The temperature is 350 K.

Because of the symmetric behavior discussed earlier, one can only move vacancies up to the middle of the device.

In this step, we can track the changes of the above quantities for a longer length of the device. The second structure has $L_D \approx 47.96$ Å, which we did not address it here. The results for this length are presented in the supplementary material. The third structure has $L_D \approx 59.03$ Å. As figure 4(a) shows, the $ZT_{max}$ occurs for a single vacancy in the middle (SV-18-24) with $\mu = 0.99$ eV. Figure 4(b) indicates that the presence of an SV in the middle of the device can increase the ZT value by 2.5 times with respect to the pristine case. It also reduces the ZT to 0.6 times of the device with $L_D \approx 34.43$ Å in the presence of SVs. The ZT also shows higher performance in comparison to the shorter case. Figure 4(c) shows a slightly downward trend in the phononic term. As the SVs get closer to the middle of the device, the ratio of the phononic term, reduces. The phononic term shows a reduction of ~ 0.8 times of the clean case, which is close to that of the shorter length of the device. In the first system, electronic terms follow each other closely, but this correlation tends to demolish in longer device, as it can be seen for electronic conductance and electronic thermal conductance for the SV-18-24.

The DVs have a slightly poor effect on the thermoelectric performance in comparison to the case of $L_D \approx 34.43$ Å. The ZT has its highest value with $\simeq 0.21$ at the chemical potential of -1.52 eV for the DV-6-8-pr. However, this double vacancy in the shorter length also gives the highest ZT. Phonon thermal conductance reduction is almost negligible compared to the shorter length, but DVs induce stronger fluctuations in electronic terms, as evidenced in figure 4(f). The ZT has a smaller value for this length in the presence of DVs.

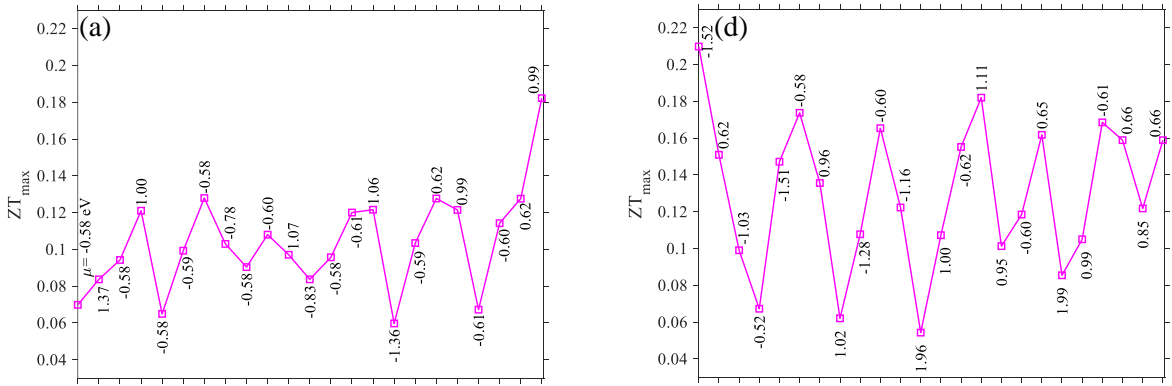

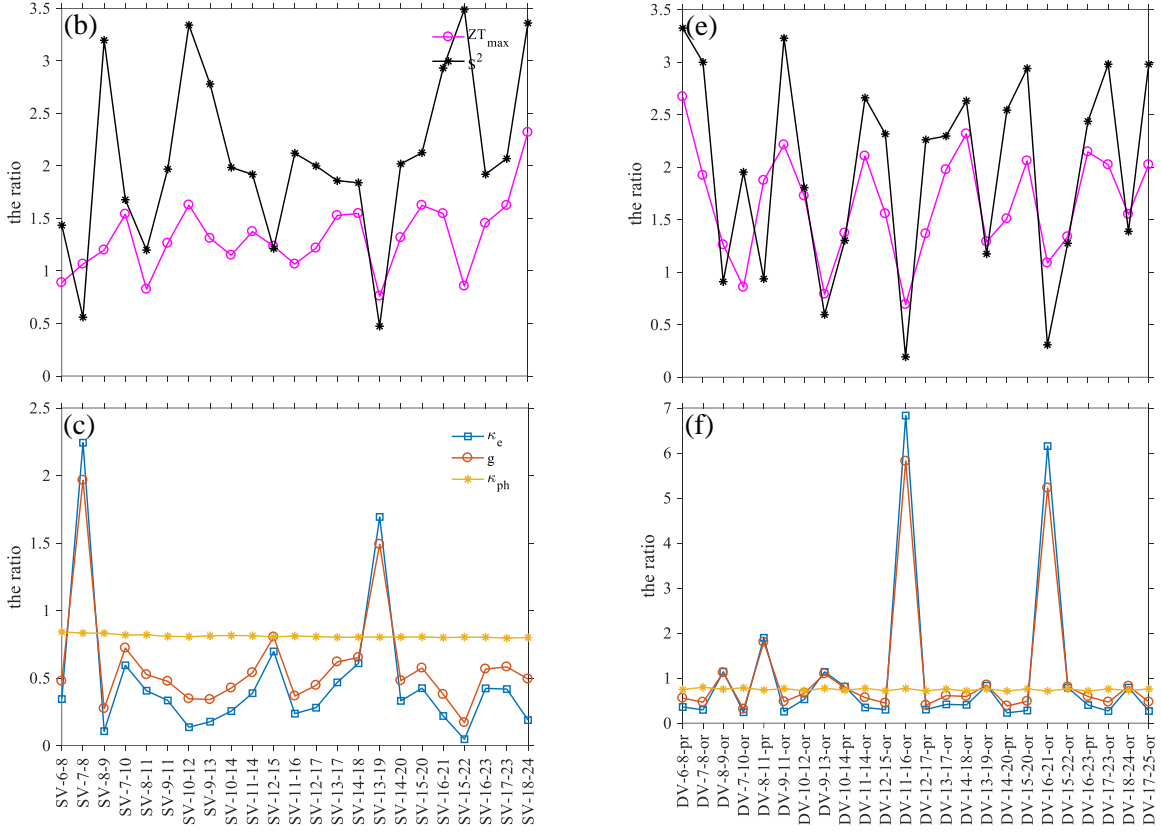

**Figure 4**. (a) Highest figure of merit variation and the corresponding $\mu$ respect to the SV locations (as described in the text), (b) *the ratio* of the Seebeck coefficient square and the maximum ZT, (c) *the ratio* of the electronic and phononic terms as a function of the SV locations for the S-shape ZGNR with a length of $\approx 59.03$ Å, (d) the $ZT_{max}$ and its corresponding chemical potential as a function of different DV locations, (e) *the ratio* of the Seebeck square and the maximum of ZT, and (f) *the ratio* of electronic and phononic terms versus the DV locations.

By comparing panels (c) and (f) of figure 4, and the same ones in figure 3, one can conclude that in the case of SVs, as the length increase, the figure of merit also rises, and $\kappa_e$ and g become more independent. For DVs, lengthening device reduces the ZT, but it decouples electronic terms. In the longer length, the fluctuation of chemical potential corresponds to the $ZT_{max}$, becomes smaller.

Vibrational local DOS (vLDOS) for the first and third lengths are shown in figure 5. Regarding figures 3(c) and (f), the lowest thermal conductance is for the SV-10-12 (or SV-13-19), which is placed on an SV with the highest vLDOS; the same is true for the DV-8-11-pr (or DV-14-20-pr) (figure 5(a)). Increasing the device length causes the system to experience the lower vLDOS (figure 5(b)). In the long devices, the higher vLDOSs are related to the edge atoms, suggesting edge defects can induce a more substantial effect on $\kappa_{ph}$.

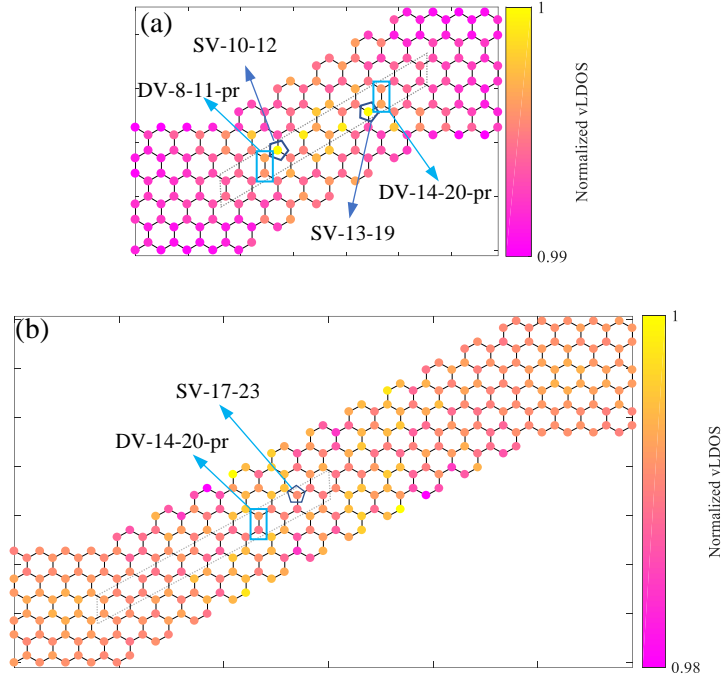

Figure 5 Phononic LDOS for a perfect device with (a) $L_D \approx 34.43$ Å, and (b) $L_D \approx 59.03$ Å. Vacancies that cause lowest thermal conductance are shown. The dotted skewed rectangle shows the zone in which vacancies are introduced.

We also studied electronic and phononic transport properties for ZGNR with the exact width of the contacts, and an AGNR similar in width to the chiral section of the device, together with the two S-shape GNR lengths studied here. As it is depicted in figure 6(a), electronic conductance for large chemical potentials (>1.5 eV) is almost close for 10-AGNR (blue line) and 12-ZGNR (green line), and two S-Shape graphene structure lengths (shorter length is indicated by red and the longer length is indicated by cyan), for the structure with a longer length, the behavior of the electronic conductance for values of $\mu$ close to zero is almost similar to that of the 10-AGNR.

The behavior for electronic thermal conductance is in a similar trend for g, as depicted in figure 6(b). Seebeck coefficient is plotted in figure 6(c) versus chemical potential, which for the shorter length, it has smaller values than the longer length. Also, the peaks between -1 and 1 for chemical potentials are opposite, which shows the change of charge carriers [11], and closeness to the AGNR behavior. The $\kappa_{\text{ph}}$ decreases by lengthening the system (check the results for $L_D \approx 47.96$ Å in the supplementary material, which confirms this trend), which is the attribution of anharmonicity of phonon modes in chiral and zigzag parts (figure 6(d)).

As the length of the chiral part increase, the AGNR characteristics become stronger. The Seebeck coefficient enhances with a bandgap [45], so one should be aware of these tradeoffs between phononic and electronic terms and the length impact when designing an efficient thermoelectric structure. Phonon mismatch between the AGNR and the ZGNR parts becomes stronger as the AGNR characteristics become dominant by lengthening the chiral section. Moreover, vibrational modes occupy a narrower frequency range in comparison to the ZGNR sections (figure 2(b) and (d)), which limits thermal conductance [43].

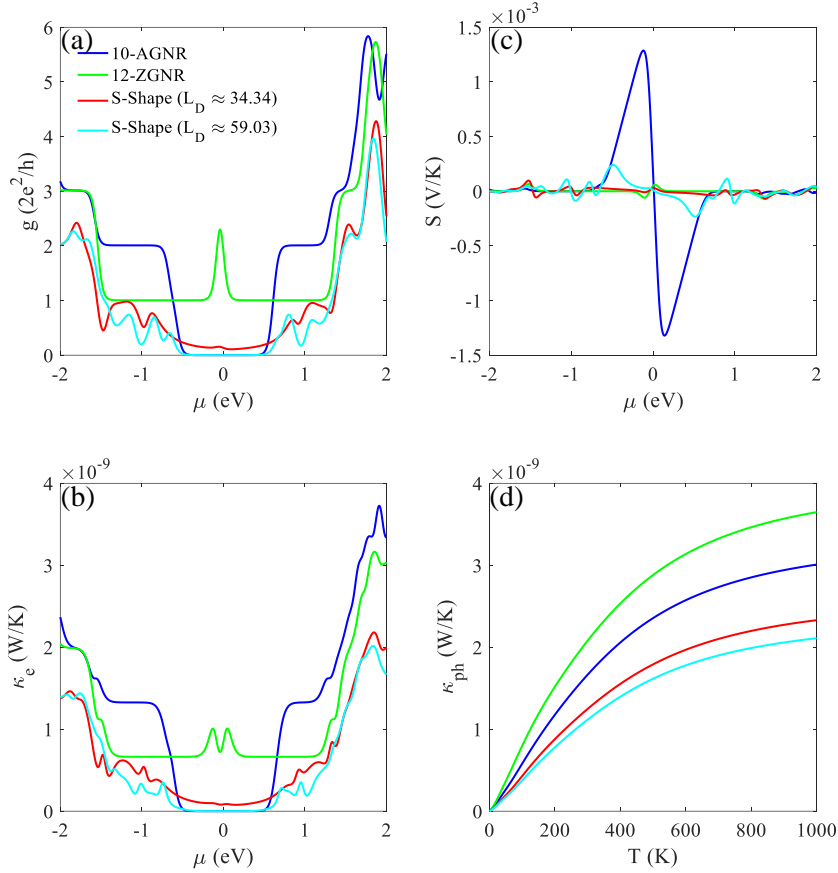

**Figure 6** (a) Electronic conductance, (b) electronic thermal conductance, and (c) Seebeck coefficient versus the chemical potential for 10-AGNR (blue line), 12-ZGNR (green line), and two different lengths of the S-Shape graphene structure (short S-shape graphene nanoribbon with the red line. The cyan line is for the longer length). (d) Phononic thermal conductance as a function of temperature.

We choose the system with $L_D \approx 59.03$ Å due to the strong increase of its ZT in the presence of SV-18-24. Here, we study the transmission coefficient for this system, both for electrons and phonons. According to figure 7(a), there is a small peak close to zero energy (cyan line) in the transmission spectrum of the system caused by the SV-18-24, which shows the close behavior of transmission coefficient of the defected S-shape GNR with the 12-ZGNR (green line). Although, the behavior of the pristine S-shape GNR is likely dominated by its chiral section (red line). The transmission spectrum of the pristine system has a semiconducting bandgap, similar to the 10-AGNR gap (blue line).

As one can see in figure 7(b), low-frequency phonons are affected more than high-frequency phonons in the pristine S-shape GNR (red line), compared to the 10-AGNR (blue line), and 12-ZGNR (green line). Besides, the transmission coefficient of the bent systems shows more suppression in the range of 0 to $1.5 \times 10^{14}$ Rad/s. This is, in general, a good change since low-frequency phonons are known to be more responsible in thermal conductance [46].

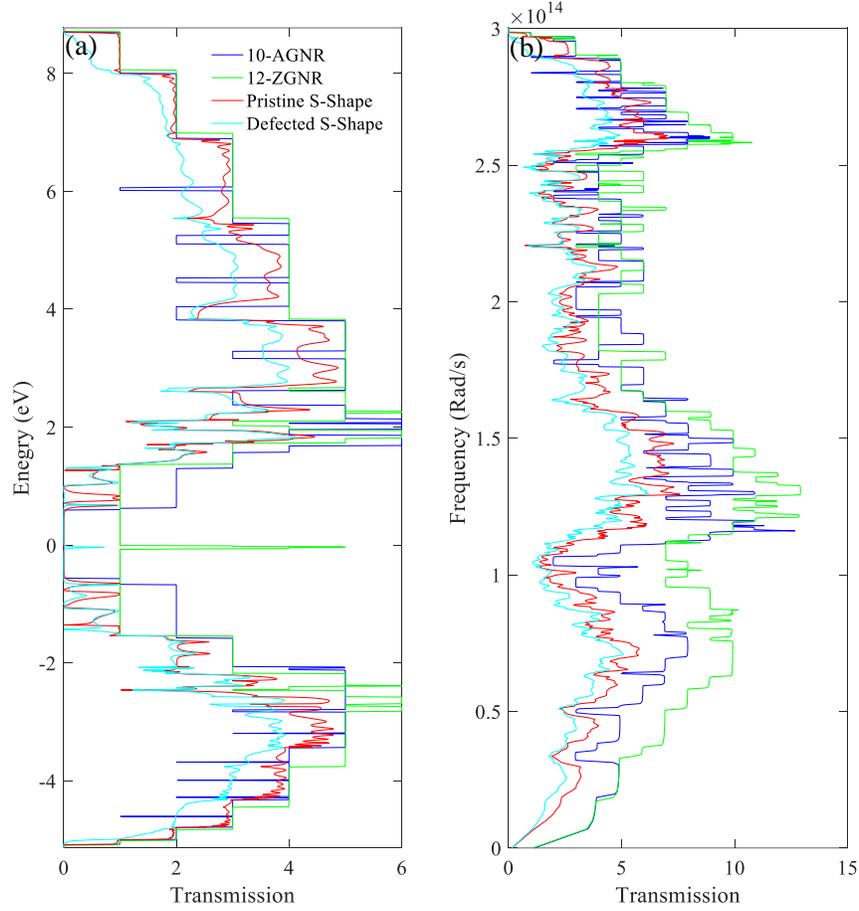

**Figure 7** (a) Transmission spectra for electrons, and (b) for phonons in the 10-AGNR (indicated by blue), 12-ZGNR (green), pristine S-shape (red), and the defected S-Shape structure with $L_D \approx 59.03$ Å (cyan).

Additionally, the impact of temperature for various chemical potentials on the ZT is plotted in figure 8 for the SV-18-24. As evidenced in this figure, the maximum ZT of 0.32 can be achieved at 1000 K for μ ≈ 0.98 eV.

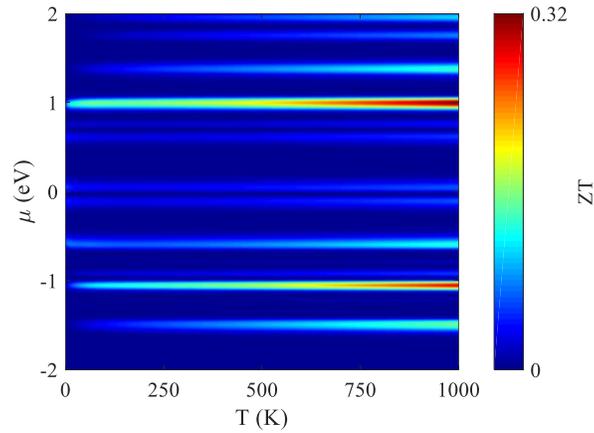

Figure 8 The ZT variation as a function of temperature and chemical potential for SV-18-24 in a system with $L_D \approx 59.03$ Å.

## 4. Conclusions

In summary, we have studied the behavior of electronic and phononic contributions in S-shape graphene structures with different lengths by including single and double vacancies in different locations. We have investigated the electronic and phononic transport properties of the system by using the NEGF method and the tight-binding approach by considering the 3NN for electronic contribution, and the force constant model with 4NN, for the phononic part.

Our numerical results show the symmetric behavior of the terms involved in the ZT, in the presence of defects respect to half of the device. The maximum of ZT is obtained for the DVs in the shorter length of the device. Also, the device with a longer length has a ZT of <0.21. By increasing the length of the chiral part of the system, the ZT can be enhanced, not only by reducing the vibrational contribution, but also by separating the electronic terms.

Single vacancies can detach the electronic thermal conductance and electronic conductance better than double vacancies; namely, the SVs can magnify the ZT up to 2.5 times in the studied structures. Detaching $\kappa_e$ and g from each other is a way to improve the ZT performance.

Lengthening the system causes the chiral section (AGNR) characteristics to become dominant.

One can tune the electronic thermal conductance and electronic conductance properties of the system by altering the parameters, such as the type and the location of the vacancy defects and the device size.

# Data availability

All data generated for this study are included in the manuscript and in the supplementary material.

# Author contributions

M.A.B carried out the simulations. M.A.B and F.K analyzed the data and prepared the manuscript. F.K. supervised the project and revised the final manuscript. All authors read and approved the final manuscript.

# Competing interests

The authors declare no competing interests.
*Corresponding author's email: khoeini@znu.ac.ir*Corresponding author's email: khoeini@znu.ac.ir

Supplementary Materials for manuscript:

# Tuning Phononic and Electronic Contributions of Thermoelectric in defected S-Shape Graphene Nanoribbons

*M.Amir Bazrafshan, Farhad Khoeini*

Here, the formalism of the NEGF is given in detail, although this method is presented in the literature, but it is in agreement with the notation of this work. The figures regarding a device length $L_D \approx 47.96$ are presented here.

The retarded Green's function reads [1]:

$$G(E) = [(E + i\eta)S_D - H_D - \Sigma_{LC}(E) - \Sigma_{RC}(E)]^{-1}, \qquad (2)$$

where $E$ is the energy, $\eta$ is an arbitrarily small positive number, and $\Sigma_{LC(RC)}$ is the self-energy for the left and right contacts, which are given by [2]:

$$\Sigma_{LC} = ((E + i\eta)S_{D-LC} - H_{D-LC})G_{0,0}^{LC}(E)((E + i\eta)S_{LC-D} - H_{LC-D}), \qquad (2a)$$

$$\Sigma_{RC} = ((E + i\eta)S_{D-RC} - H_{D-RC})G_{2,2}^{RC}(E)((E + i\eta)S_{RC-D} - H_{RC-D}), \qquad (2b)$$

where $G_{LC(RC)}(E)$ is the isolated left (right) contact surface Green's function. Retarded surface Green's functions are also given by:

$$G_{0,0}^{LC}(E) = \left((E + i\eta)S_{0,0}^{LC} - H_{0,0}^{LC} - ((E + i\eta)S_{-1,0}^{LC} - H_{-1,0}^{LC})\tilde{\Lambda}\right)^{-1}, \qquad (3a)$$

$$G_{2,2}^{RC}(E) = \left((E + i\eta)S_{2,2}^{RC} - H_{2,2}^{RC} - ((E + i\eta)S_{2,3}^{RC} - H_{2,3}^{RC})\Lambda\right)^{-1}, \qquad (3b)$$

where $\Lambda$ and $\tilde{\Lambda}$ are transfer matrices. Transfer matrices for the contacts can be obtained as:

$$\tilde{\Lambda} = \tilde{t}_0 + t_0\tilde{t}_1 + t_0 t_1\tilde{t}_2 + \cdots\cdots\cdots + t_0 t_1 t_2 \cdots \tilde{t}_n\ ; \qquad (4a)$$

$$\Lambda = t_0 + \tilde{t}_0 t_1 + \tilde{t}_0\tilde{t}_1 t_2 + \cdots\cdots\cdots + \tilde{t}_0\tilde{t}_1\tilde{t}_2 \cdots t_n\ , \qquad (4b)$$

where $t_0, \tilde{t}_0, t_i$ and $\tilde{t}_i$ for the left contact can be given as:

$$t_0 = \left((E + i\eta)S_{0,0}^{LC} - H_{0,0}^{LC}\right)^{-1}\left((E + i\eta)S_{0,-1}^{LC} - H_{0,-1}^{LC}\right), \qquad (5a)$$

$$\tilde{t}_0 = \left((E + i\eta)S_{0,0}^{LC} - H_{0,0}^{LC}\right)^{-1}\left((E + i\eta)S_{-1,0}^{LC} - H_{-1,0}^{LC}\right),$$

$$t_i = (I - t_{i-1}\tilde{t}_{i-1} - \tilde{t}_{i-1}t_{i-1})^{-1}t_{i-1}^2\ , \qquad (5b)$$

$$\tilde{t}_i = (I - t_{i-1}\tilde{t}_{i-1} - \tilde{t}_{i-1}t_{i-1})^{-1}\tilde{t}_{i-1}^2\ , \qquad (5c)$$

$$(5d)$$

with I as identity matrix. With a same procedure, one can form the transfer matrices for the right contact. By considering $S_D = I$ and other overlap matrices are zero, and also replacing $E$ by $\omega^2$ this formalism can be used to predict phonon transport [3]. Since every atom has 3 spatial degrees of freedom, every element of the Hamiltonian matrices, is a $3 \times 3$ block matrix in the dynamical matrix.

The spectral density operator is given by:

$$\Gamma_{LC(RC)}(E) = i[\Sigma_{LC(RC)}(E) - \Sigma_{LC(RC)}(E)^\dagger]. \qquad (7)$$

The transmission probability for the electron (phonon) can be evaluated by:

$$T_{e(ph)}(E) = \text{Trace}[\Gamma_{LC}(E)G(E)\Gamma_{RC}(E)G(E)^\dagger]. \qquad (8)$$

Furthermore, the DOS and local DOS can be obtained using:

$$\text{DOS}(E) = -\frac{1}{\pi}\text{Im}\left(\text{Tr}(G(E))\right), \qquad (9)$$

$$\text{LDOS}(E) = -\frac{1}{\pi}\text{Im}\left(\text{DIAG}(G(E))\right). \qquad (10)$$

For phonons, DOS and LDOS are changed to vDOS and vLDOS. As mentioned earlier, for phonon $E$ is replaced with $\omega^2$. For each atom, there are 3 values for vLDOS, in this work we summed these values.

Results for $L_D \approx 47.96$ Å are shown in Figure S 1.

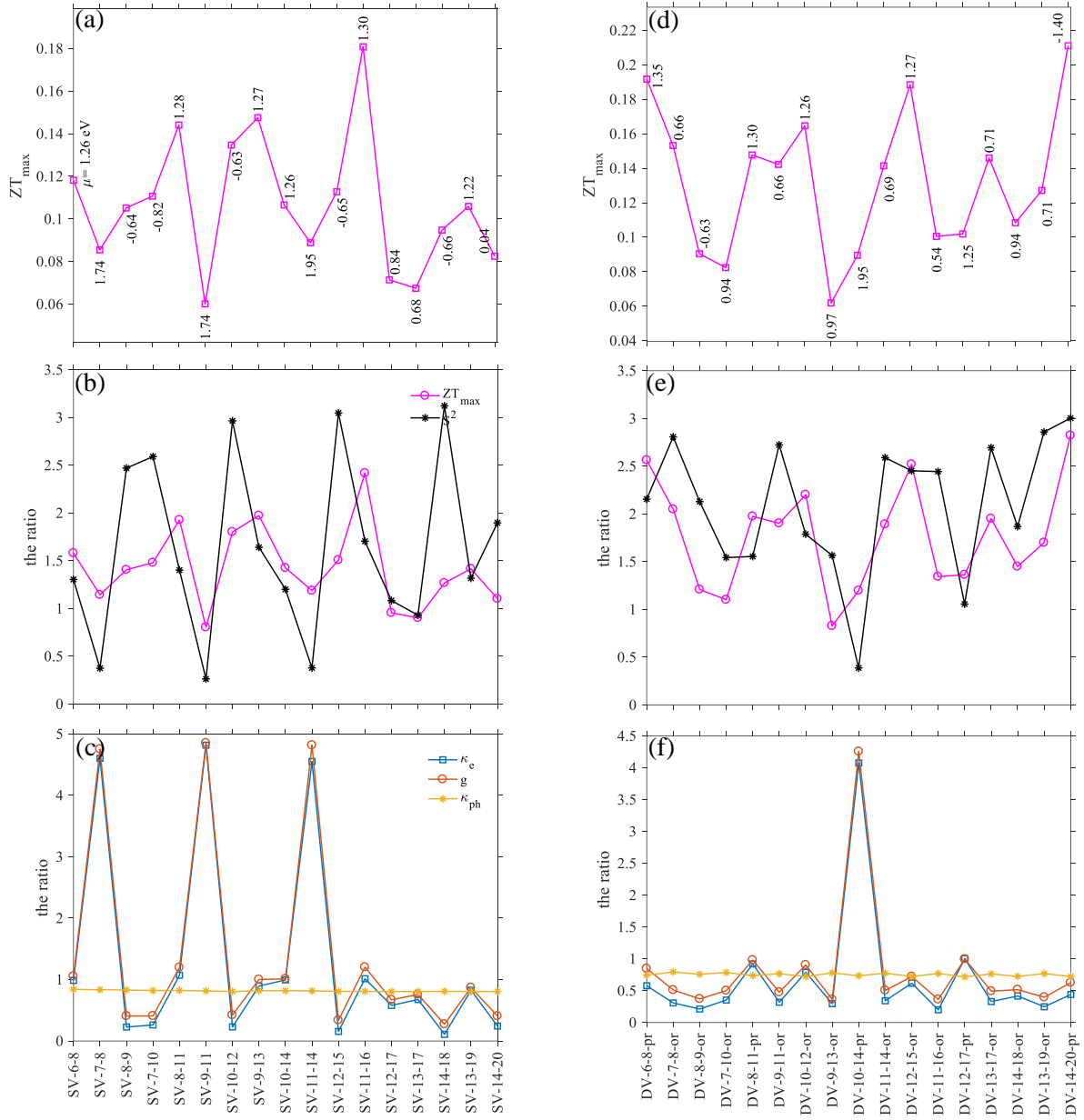

Figure S 1. (a) Maximum figure of merit for different SV locations together with the corresponding $\mu$, (b) the ratio of $S^2$ and $ZT_{max}$, and (c) electronic and phononic terms as a function of various SV locations, (d) the ratio of $ZT_{max}$ and its chemical potential as a function of different DV locations, (e) the ratio of the Seebeck square and maximum of ZT, and (f) *the ratio* of electronic and phononic terms vs DV locations. The device length is $L_D \approx 47.96$ Å.

The dependence of the ZT to the chemical potential and the temperature is plotted in the manuscript for SV-18-24. However, here the same plot for the pristine system (Figure S 2) and with a DV, specifically the one that induces significant increase in ZT, DV-6-8-pr (Figure S 3), are also included.

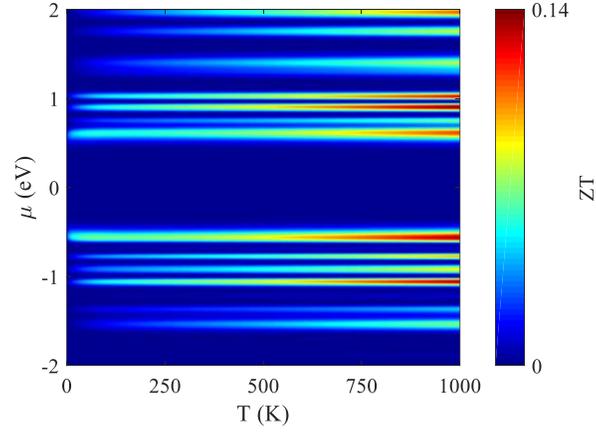

Figure S 2. The ZT as a function of chemical potential and temperature for the perfect system with $L_D \approx 59.03$ Å.

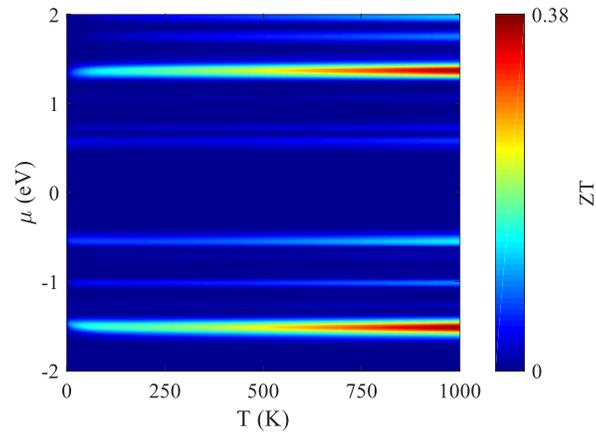

Figure S 3. The ZT as a function of chemical potential and temperature in the presence of the DV-6-8-pr for the system with $L_D \approx 59.03$ Å.

In Figure S 4, we show the result presented in figure 6 for a system with $L_D \approx 47.96$ Å.

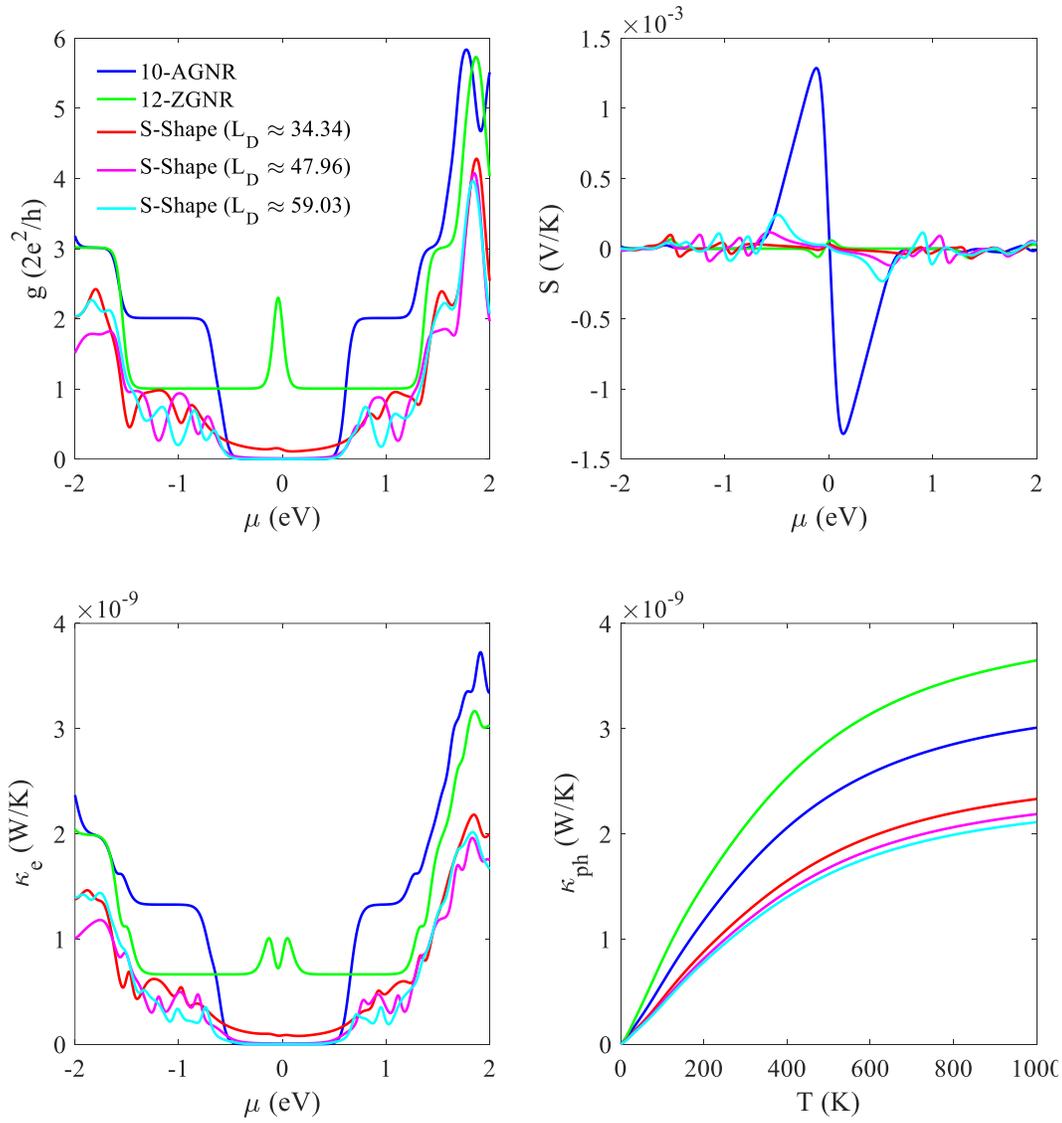

Figure S 4. Comparison between the pristine case of all studied structures with the pristine GNRs in the S-Shape structure. The unit of length is angstrom.

A series of figures related to the vLDOS and LDOS of systems are presented in the following based on increasing lengths.

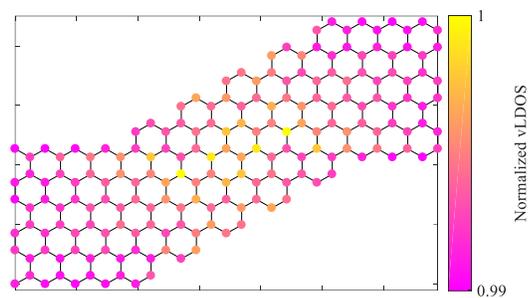

Figure S 5. The system with $L_D \approx 34.43$ Å. The figure shows that the high vLDOS atoms are almost far from the edges.

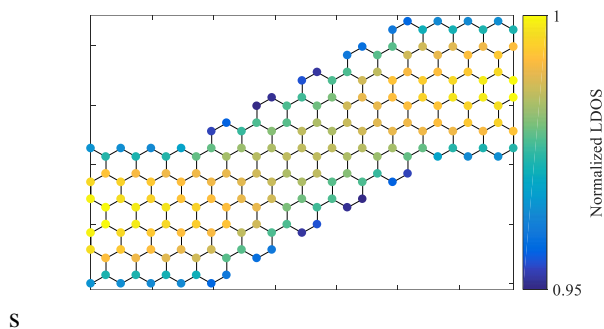

s

Figure S 6. The system with $L_D \approx 34.43$ Å. The figure shows that the high LDOS atoms are almost far from center (purple dashed line in figure 1).

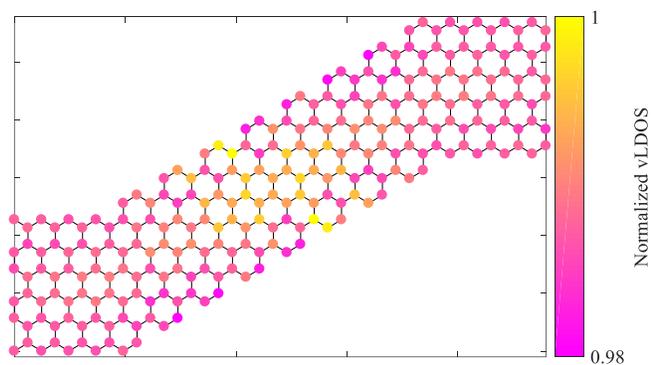

Figure S 7. The system with $L_D \approx 47.96$ Å. The atoms with high vLDOS are not concentrated in the middle of the chiral part.

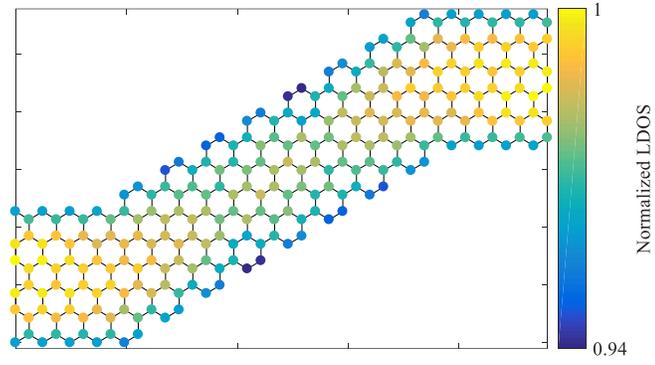

Figure S 8. The system with $L_D \approx 47.96$ Å. The atoms that are located in the middle, have higher LDOS values from their neighbors.

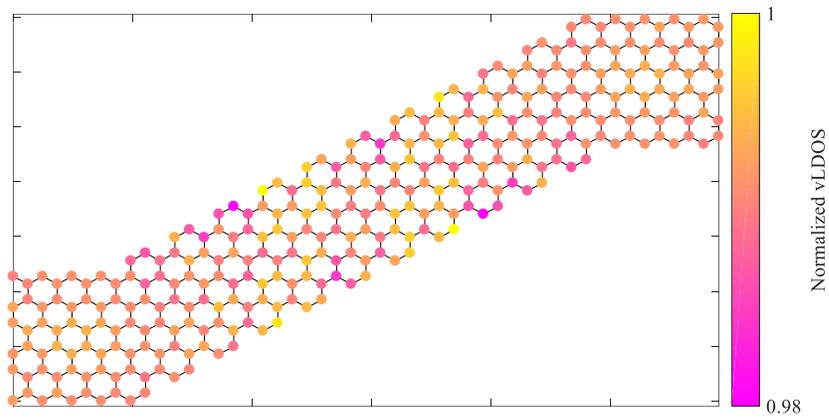

Figure S 9. The system with $L_D \approx 59.03$ Å. The high vLDOS atoms are almost on the edges.

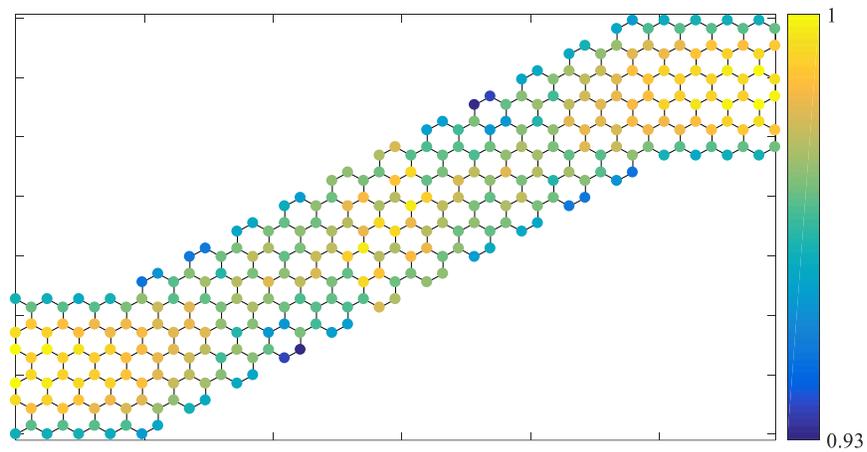

Figure S 10. The system with $L_D \approx 59.03$ Å. The atoms with high LDOS are appeared in the middle of the structure.

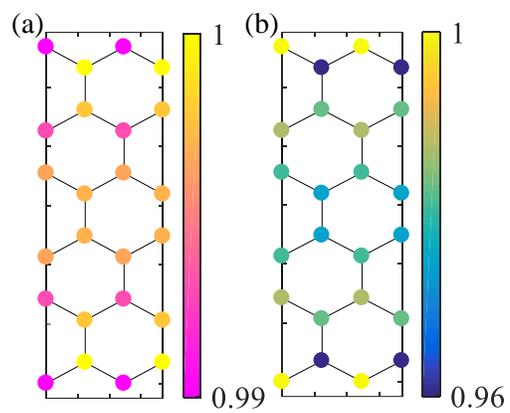

Figure S 11. (a) The vLDOS, and (b) LDOS for 12-ZGNR.

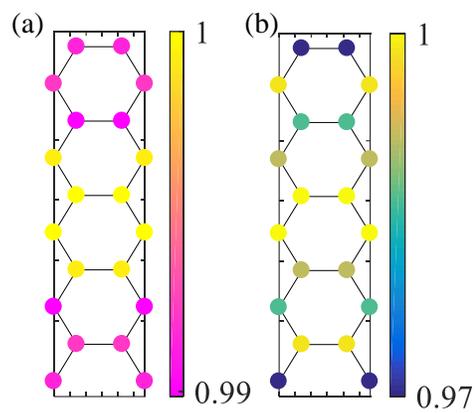

Figure S 12. (a) The vLDOS, and (b) LDOS for 10-AGNR.